\documentclass[journal=jacsat,manuscript=article]{achemso}
\usepackage[version=3]{mhchem}

\setkeys{acs}{etalmode=truncate,maxauthors=10}  

\usepackage{graphicx}
\usepackage{dcolumn}
\usepackage{comment}
\usepackage{bm}
\usepackage{xcolor}
\usepackage{booktabs}
\usepackage{gensymb}
\usepackage{mathtools}
\usepackage{braket}
\usepackage{multirow}
\usepackage[normalem]{ulem} 
\SectionNumbersOn

\author{Prakash Mishra}
\affiliation[The University of Texas at El Paso]
{Computational Science Program, University of Texas at El Paso, El Paso, Texas 79968, USA}
\author{Yoh Yamamoto}
\affiliation[Unknown University]
{Department of Physics, University of Texas at El Paso, El Paso, TX, 79968, USA}
\author{Po-Hao Chang}
\affiliation[Unknown University]
{Department of Physics, University of Texas at El Paso, El Paso, TX, 79968, USA}
\author{Duyen B. Nguyen}
\affiliation[Unknown University]
{Physics Department and Science of Advanced Materials Program, Central Michigan University, Mt. Pleasant, Michigan 48859, USA}
\author{Juan E. Peralta}
\affiliation[Unknown University]
{Physics Department and Science of Advanced Materials Program, Central Michigan University, Mt. Pleasant, Michigan 48859, USA}
\author{Tunna Baruah}
\affiliation[Unknown University]
{Computational Science Program, University of Texas at El Paso, El Paso, Texas 79968, USA}
\alsoaffiliation[Second University]
{Department of Physics, University of Texas at El Paso, El Paso, TX, 79968, USA}
\author{Rajendra R. Zope}
\email{rzope@utep.edu}
\affiliation[Unknown University]
{Computational Science Program, University of Texas at El Paso, El Paso, Texas 79968, USA}
\alsoaffiliation[Second University]
{Department of Physics, University of Texas at El Paso, El Paso, TX, 79968, USA}

\title[title]
  {Study of Self-Interaction Errors in Density-Functional Calculations  of Magnetic Exchange Coupling Constants Using Three Self-Interaction-Correction Methods}

\begin{document}

\begin{abstract} 
  
  We examine the role of self-interaction errors (SIE) removal on the evaluation of magnetic exchange coupling constants. In particular we analyze the effect of scaling down the self-interaction-correction (SIC) for three {\em non-empirical}  density functional approximations (DFAs) namely, the local spin density approximation, the Perdew-Burke-Ernzerhof generalized gradient approximation, and recent SCAN family of meta-GGA functionals. To this end, we employ three one-electron SIC methods: Perdew-Zunger [Perdew, J. P.; Zunger, A. \textit{Phys. Rev. B}, {\bf 1981}, \textit{23}, 5048]
   SIC, the orbitalwise scaled SIC method [Vydrov, O. A. \textit{et al.}, \textit{J. Chem. Phys.} {\bf 2006,} \textit{124}, 094108],
   and the recent {local} scaling method [Zope, R. R. \textit{et al.},
  \textit{J. Chem. Phys.} {\bf 2019}, \textit{151}, 214108].
   We compute the magnetic exchange coupling constants using the spin projection and non projection 
  approaches for  sets of molecules 
   composed of dinuclear and polynuclear H--He models, organic radical molecules, and chlorocuprate,
  and compare these results against accurate theories and experiment. 
   Our results show that  for the systems that mainly consist of single electron
regions, PZSIC performs well but for more complex organic systems and the chlorcuprates,
an overcorrecting tendency of PZSIC combined with the DFAs utilized in this work is more pronounced, and in such cases 
LSIC with kinetic energy density ratio performs better than PZSIC.
  Analysis of the results in terms of SIC corrections to the density and to the total 
  energy shows that both density and energy correction are required to obtain 
  an improved prediction of magnetic exchange couplings.

\end{abstract}

\maketitle

\section{\label{sec:introduction}Introduction}
 The characterization of the magnetic properties of materials is crucial  for applications 
 such as memory storage devices, spintronics, quantum computing, and magnetic sensors. 
 In many cases, the magnetic behavior arises from interactions between localized electrons  
 typically described using model spin Hamiltonians,  including  spin-spin exchange and dipolar 
 interactions, and magnetic anisotropy contributions.
 The accurate theoretical description of magnetic properties such as exchange interactions
or spin levels by means of  electronic structure calculations is especially challenging
since the energy spacing  involved can be much smaller than the typical electronic excitation 
energies. 
In particular, the calculation of magnetic exchange couplings ($J$) in transition metal 
complexes  have been the focus of attention for  density functional theory methods.\cite{doi:10.1063/1.2145878,doi:10.1063/1.1999631,JOSHI2020114194,MELO2013110,doi:10.1021/ct4007376,https://doi.org/10.1002/pssb.200541490} 
In this respect, some approximations can be successful in predicting $J$ couplings for certain type of complexes but fail for others.
Hybrid density functional approximations, in their  different flavors, and recently the strongly constrained appropriately
 normed (SCAN) meta-generalized-gradient-approximation functional (meta-GGA), have shown some promise for predicting magnetic properties, including  $J$ couplings.
\cite{doi:10.1021/ja961199b, 
doi:10.1063/1.4752398, 
Pantazis_2019,yamamoto2020assessing}

Experimentally, coupling parameters are obtained by fitting susceptibility data. For complex structures 
with multiple exchange pathways, this fitting can become  very challenging or even impossible.  
Other
experimental techniques, such as electron paramagnetic resonance (EPR), electron spin resonance (ESR),
or inelastic neutron scattering (INS) are also used to determine exchange couplings, but they  present 
the same type of shortcomings. Thus, for large, multi-center magnetic complexes, computational methods
that can predict $J$ couplings  are the only choice to determine these constants.
Electronic structure methods are routinely used to  determine the coupling parameter by 
mapping  energies of different  spin states of spin Hamiltonians.\cite{doi:10.1021/ja00850a018,noodleman1981valence,NOODLEMAN1992423,NOODLEMAN1995199}
Among electronic structure methods, density functional theory (DFT)\cite{PhysRev.136.B864,kohn1965self} 
is widely used because it provides a reasonable balance between accuracy and computational efficiency. 
For the case of multi-center transition metal complexes,  computational efficiency becomes crucial, 
and therefore methods based on DFT are, many times, the only alternative. 
Unfortunately, approximations to the exchange-correlation functionals used in density functional 
calculations suffer from self-interaction error (SIE).\cite{PhysRevB.23.5048,zunger1980self}
This error emerges when the approximate exchange-correlation energy does not completely
cancel out the Coulomb interaction of an electron with itself for one-electron systems.  
It is well documented  that the presence of SIE hinders the accurate description of many
properties 
involving chemical reactions, dissociation, charge transfer, and magnetism.\cite{Ziegler2002,doi:10.1063/1.1630017,doi:10.1021/ja039556n,PEDERSON2015153,doi:10.1063/1.2085171}
The SIE in standard approximate functionals results in excessive electron delocalization, causing an error in predicted properties that is known as delocalization 
error.\cite{zhang1998challenge,ruzsinszky2006spurious,mori2006many}
The Heisenberg exchange coupling parameters  are very sensitive in sign and magnitude to the overlap between orbitals, both involving spin centers and  bridging 
ligands,\cite{anderson1950antiferromagnetism,kollmar1993ferromagnetic,hart1992ab} and 
hence  the delocalization error, which affects the spin-density at the metal centers, gives rise to large changes in the sign and strength of the calculated coupling  parameters $J$.

One approach to mitigate SIE in DFAs is to use hybrid functionals such as the hybrid version of the Perdew-Burke-Ernzerhoff (PBEh)\cite{PBEh} or the popular B3LYP (Becke's 3-parameter for exchange and Lee-Yang-Parr for correlation).\cite{B3LYP}  
Over the years several groups have documented the performance of various DFAs 
for $J$ couplings.\cite{doi:10.1063/1.2085171,illas2004extent,moreira2007restricted,peralta2010magnetic,phillips2011role,valero2008performance,Pantazis_2019,
doi:10.1021/acs.jpca.7b12663} 
The general conclusions of these studies suggest that  there is no single approach that 
performs best across different types of couplings involving different transition metals, and hence the predictive power of a
particular DFA approach is limited. In view of this, other approaches based on DFT 
that can reduce SIE are specially attractive candidates for the prediction of $J$ couplings.

Perdew and Zunger, in 1981, proposed a scheme (PZSIC) to explicitly remove the SIE on an orbital-by-orbital 
basis,\cite{PhysRevB.23.5048} making the approximate energy functional exact for any one-electron density (see Section~\ref{sec:theory}). 
Their work extends the earlier self-interaction scheme for $X$-$\alpha$ exchange due to Lindgren\cite{lindgren1971statistical} 
to the local density functional approximation.
Ruiz \textit{et al.}\cite{doi:10.1063/1.2085171} studied the effect of SIE on exchange couplings using PZSIC to 
determine the $J$  coupling in a simple model system, H--He--H. 
In recent years, the development of the Fermi-L\"owdin orbital self-interaction correction (FLOSIC) methodology\cite{doi:10.1063/1.4869581,yang2017full,doi:10.1063/1.4907592,PEDERSON2015153,doi:10.1063/1.4996498} and software\cite{FLOSICcode,FLOSICcodep,yamamoto2020assessing,doi:10.1063/5.0056561,PhysRevA.103.042811,schwalbe2020pyflosic,https://doi.org/10.1002/jcc.26168,doi:10.1063/5.0031341,doi:10.1063/5.0004738} improved  the accessibility of SI-free calculations.
FLOSIC has been applied to study various molecular properties and showed improvement over LSDA and GGA for most cases where SIE dominates the errors. 
\cite{doi:10.1063/1.4996498,PhysRevA.100.012505,doi:10.1063/1.5087065,doi:10.1063/1.5120532,Sharkas11283,joshi2018fermi,C9CP06106A,Jackson_2019,doi:10.1063/1.5125205,doi:10.1002/jcc.25767,doi:10.1021/acs.jctc.8b00344,waterpolarizability,doi:10.1021/acs.jpca.8b09940,doi:10.1063/1.4947042,kao2017role,doi:10.1063/5.0041265}
Joshi \textit{et al.}\cite{joshi2018fermi} used FLOSIC with LSDA  to investigate the effect of SIE on more realistic systems than the simple model
H--He--H,  and found that removing the SIE corrects $J$ couplings towards more accurate methods such as 
 hybrid density functionals and  wave function based methods, in line with the earlier study 
 of Ruiz \textit{et al.}\cite{doi:10.1063/1.2085171}

 It is well known that  that PZSIC tends to overcorrect, particularly for the equilibrium properties,
 resulting in errors of opposite sign to those from semilocal functionals.\cite{vydrov2006scaling,perdew2015paradox} 
 Recently, Zope \textit{et al.}\cite{doi:10.1063/5.0010375} 
 introduced the locally scaled SIC 
 (LSIC), which uses a pointwise iso-orbital indicator to identify the one electron self-interaction regions
 in many-electron system (see Section~\ref{sec: local-scaling}) and to scale down the SIC in the many-electron regions. 
 LSIC  works well for both equilibrium properties as well as for properties that involve a stretched bond, and 
 provides an improved performance with respect to standard SIC for a wide 
 range of electronic structure properties.
 \cite{doi:10.1063/1.5129533, waterpolarizability, doi:10.1063/5.0041265, doi:10.1063/5.0041646, akter2021well} 
 Motivated by the success of LSIC, in this work we explore the potential of 
 locally scaled SIC methods for the prediction of magnetic exchange coupling constants $J$. 
 We also perform more comprehensive study of coupling constants by applying three 
 different SIC methods to the three levels of exchange-correlation approximations 
 that include the most recent r$^2$SCAN meta-GGA functional.
 The two SIC methods besides  the well known PZSIC method\cite{PhysRevB.23.5048} are
 the orbitalwise scaled down  SIC method of Vyrdov {\it et al.} \cite{vydrov2006scaling} and 
 the recent LSIC method of Zope and coworkers.\cite{doi:10.1063/1.5129533} 
 Results of various SIC methods  for the three different class of systems 
 are compared and analyzed.

\section{Theory}\label{sec:theory}
\subsection{Magnetic exchange coupling constants}

 Isotropic spin-spin interaction between two magnetic centers $A$ and $B$  are usually described by means of the Heisenberg-Dirac-Van Vleck (HDVV) spin Hamiltonian,\cite{heisenberg1985theorie}
 \begin{equation}\label{eq:HDVV}
 H_{HDVV}=-{\sum_{A,B} J_{AB} \vec{S}_A \cdot \vec{S}_B}\,, 
 \end{equation}
 where $\vec{S}_A$ ($\vec{S}_B$) is the spin operator for the site $A$ ($B$), and $J_{AB}$ is the exchange coupling constant.
 The magnitude and sign of $J_{AB}$ is related to  the strength and the nature of the coupling (ferromagnetic or antiferromagnetic).
To determine $J$ couplings from electronic structure calculations, it is necessary to map the energy spectrum of the HDVV Hamiltonian to the electronic structure energy spectrum. To illustrate this, consider a two spin-1/2 case. The energy difference between the singlet  or low-spin state $\ket{\text{LS}}=\ket{S\!\!=\!\!0}$ and any high-spin or triplet state $\ket{\text{HS}}=\ket{S\!\!=\!\!1}$ from the HDVV Hamiltonian is
$E_\text{HS}-E_\text{LS}=J_{AB}$
. However, for single-reference methods such as those based on DFT, the singlet state is not accessible. In these cases, one can resort to evaluating the energy difference between the broken-symmetry low-spin state $\ket{\text{BS}}=\ket{\uparrow\downarrow}$ and the 
high-spin (HS) state $\ket{\text{HS}}=\ket{\uparrow\uparrow}$, which for the two spin-1/2 case is  
$E_\text{BS}-E_\text{HS}=J_{AB}/2$. This approach is exact for perfectly localized magnetic centers, and it can be generalized to two spins $S_A$ and $S_B$ to extract the exchange coupling as 
\begin{equation}\label{eq:Jsp}
    J_\text{SP} = \frac{E_\text{BS}-E_\text{HS}}{2 S_A S_B},
\end{equation}
where SP indicates spin-projected. 
This approach was introduced by Noodleman\cite{noodleman1981valence} and is widely used in the literature. For cases where there is significant overlap between the magnetic orbitals, one can use the $\ket{\text{HS}}$ and $\ket{\text{LS}}$ states to evaluate the expectation values of the HDVV Hamiltonian. 
This gives place to the  non-projected (NP) approach of Ruiz \textit{et al.},\cite{RuizJCC1999}   where 
\begin{equation}\label{eq:Jnp}
    J_\text{NP}=\frac{E_\text{LS}-E_\text{HS}}{2S_A S_B +S_B}\,.
\end{equation}
For single-reference states, this approach is exact for fully delocalized spins. 
It should also be mentioned, although it is not employed in this work, Yamaguchi's approach, which interpolates between both NP and SP approaches,\cite{nishino1997theoretical}

\begin{equation}\label{eq:JY}
    J_\text{Y}=\frac{2 (E_\text{BS}-E_\text{HS})}{\braket{S^2}_\text{HS}-\braket{S^2}_\text{BS}}\,.
\end{equation}

The relation between static correlation in the DFAs and SIE has been discussed in the literature\cite{Polo2003,POLO2002469,doi:10.1080/00268970110111788,polo2002some,PhysRevA.51.4531}. It has been argued that DFA calculations can contain spurious static correlation which causes the BS energy to correspond to the singlet state energy. This may be different when SIE is removed. Since the NP approach approximates the singlet energy with the BS energy, the bare DFA (SIE-uncorrected) NP approach can give better estimates of coupling constants while in the case of the SP approach this will result in double counting of the static correlation contributions\cite{doi:10.1063/1.2178793,doi:10.1063/1.2178791}.

\subsection{PZSIC and FLOSIC}
In the PZSIC\cite{PhysRevB.23.5048} method,
the one-electron SIE is removed from the energy functional by  
 subtracting the orbitalwise self-Hartree and self-exchange, and correlation as follows,
\begin{equation}\label{eq:pzsic}
    E^{PZSIC}[\rho_\uparrow,\rho_\downarrow] = E^{DFA}[\rho_\uparrow,\rho_\downarrow] - \sum_{i\sigma}^{occ}\{ U[\rho_{i\sigma}] + E_{XC}^{DFA}[\rho_{i\sigma},0] \}\,,
\end{equation}
where $i$ and $\sigma$ are the orbital and spin indices, and $\rho_{i\sigma}$ is the density for orbital $i\sigma$. 
In the traditional PZSIC approach, one solves the  localization equations, \cite{doi:10.1063/1.454104} 
to determine the set of localized orbitals that minimize the PZSIC total energy. 
The computational cost of solving localizations equation can be a limiting factor 
for practical applications due to its poor scaling.\cite{doi:10.1063/1.4869581} 
In recent years, the FLOSIC approach\cite{doi:10.1063/1.4869581,
doi:10.1063/1.4907592,PEDERSON2015153,yang2017full} for solving the PZSIC problem was proposed. 
FLOSIC is a size-extensive implementation of the PZSIC method, and its computational efficiency affords the  study of large systems.
In FLOSIC, using $3N$ positions in space (called Fermi-orbital descriptors, or FODs),  $\bm{a}_j$, the set of KS orbitals $\psi_i$ is transformed to a set of Fermi orbitals $F_j$ as follows\cite{Luken1982},
\begin{equation}\label{eq:fod}
    F_j(\bm{r}) = \frac{\sum_i \psi_i(\bm{a}_j) \psi_i(\bm{r})}{\sqrt{\rho (\bm{a}_j)}}\,.
\end{equation}
The resulting Fermi orbitals are normalized but not necessarily orthogonal.  The L\"owdin orthogonalization scheme\cite{lowdin1950non} is used to form a set of orthonormal Fermi-L\"owdin orbitals (FLOs). The  FLOs are then used to solve Eq. (\ref{eq:pzsic}). 
Since FLOs depend on the FODs, minimizing the PZ energy with respect to the choice of localized orbitals is equivalent to finding the set of FODs that minimize the energy. 
Thus, the FLOSIC method requires optimizing the $3N$  $\bm{a}_j$ parameters instead of $N^2$ parameters, 
as in the localization equation approach. For more details on the FLOSIC methodology, we refer the reader to 
Refs.~\citenum{doi:10.1063/1.4907592,PEDERSON2015153,yang2017full, doi:10.1063/1.4869581}.

\subsection{Orbitalwise scaled SIC} 
PZSIC eliminates one-electron SIE and improves properties  for which DFAs perform poorly 
due to SIE, but it also worsens  the equilibrium properties where SI-uncorrected DFAs tend to perform well.
This paradoxical behavior of PZSIC\cite{perdew2015paradox} is due to an over-correcting 
tendency of PZSIC in the many-electron region.
In the orbital scaling approach (OSIC) of Vydrov \textit{et.al},\cite{vydrov2006scaling} an orbitalwise
scaling factor $X_{i\sigma}$ is introduced in Eq.~(\ref{eq:pzsic}) to reduce the over-correction as follows,
\begin{equation}\label{eq:osic}
    E^{OSIC}[\rho_\uparrow,\rho_\downarrow] = E^{DFA}[\rho_\uparrow,\rho_\downarrow] - \sum_{i\sigma}^{occ} X_{i\sigma} \{ U[\rho_{i\sigma}] + E_{XC}^{DFA}[\rho_{i\sigma},0] \}\, .
\end{equation}
A quantity such as $ \int \rho_i(\vec{r})\tau^W(\vec{r})/\tau(\vec{r}) d\vec{r}$ may be used as the orbital scaling factor $X_{i\sigma}$ where $\tau$   and $\tau^W$  are the 
 non-interacting (positive) and von Weizs\"acker kinetic energy densities, respectively.
Alternatively, $\int \rho_i(\vec{r})^2/\rho(\vec{r}) d\vec{r}$ can also be used for $X_{i\sigma}.$\cite{doi:10.1063/1.2204599}
OSIC provides improvement over PZSIC for atomization energies but it reintroduces some errors 
corrected by the PZSIC scheme. For example, unlike in PZSIC, 
the dissociation of heteronuclear molecules in OSIC showed spurious fractional charge dissociation.\cite{doi:10.1063/1.2566637}

Unlike the OSIC and PZSIC methods, the recent LSIC method\cite{doi:10.1063/1.5129533} 
described in the next section 
showed promising results for both the equilibrium properties as well as stretched 
bond properties when used in combination with the LSDA functional.
\subsection{Local-scaling SIC methods} \label{sec: local-scaling}

In the LSIC approach introduced by Zope \textit{et al.}\cite{doi:10.1063/1.5129533}, 
the magnitude of the SIC correction is determined  locally (pointwise) in space using  
a local scaling factor $f(\vec{r})$, also known as iso-orbital indicator. 
Thus, the SIC energy in the  the LSIC method is given by
\begin{align}\label{eq:LSIC-DFA}
    E_{XC}^{LSIC-DFA}[\rho_{\uparrow},\rho_{\downarrow}] = E_{XC}^{DFA}[\rho_{\uparrow},\rho_{\downarrow}]
                       - \sum_{i\sigma}^{occ}
     \left \{ U^{LSIC}[\rho_{i\sigma}]
     + E_{XC}^{LSIC}[\rho_{i\sigma},0] \right \}\,,
\end{align}
where 
\begin{align}\label{eq:lsiccoul}
     U^{LSIC}[\rho_{i\sigma}] = 
       \frac{1}{2} \int d\vec{r}\,
       f_{\sigma}(\vec{r}) \,
       \rho_{i\sigma}(\vec{r}) 
       \int d\vec{r'}\, \frac{\rho_{i\sigma}(\vec{r'})}{\vert \vec{r}-\vec{r'}\vert}
\end{align}
and
\begin{align}\label{eq:lsicxc} 
     E_{XC}^{LSIC}[\rho_{i\sigma},0]
     =  \int d\vec{r}\,  f_{\sigma}(\vec{r})
     \rho_{i\sigma} (\vec{r}) \epsilon_{XC}^{DFA}([\rho_{i\sigma},0],\vec{r}) 
\end{align}
are the scaled-down self-Hartree and self-exchange-correlation energies.
In the original LSIC work,\cite{doi:10.1063/1.5129533} 
Zope \textit{et al.} used $z_\sigma(\vec{r}) = \tau_{\sigma}^W(\vec{r})/\tau_\sigma(\vec{r})$ 
for $f_\sigma(\vec{r})$. 
This iso-orbital indicator interpolates between the single-orbital regions ($z=1$) and the uniform density regions ($z=0$),  and it is  
used as a weight for integrating the SIC energy densities accordingly. 
$z_\sigma$ is not necessarily the only choice with LSIC, and other local scaling factors can be used and sometimes preferred\cite{romero2021local}.
Hereafter, we refer to  LSIC with $z_\sigma(\vec{r}) = \tau_{\sigma}^W(\vec{r})/\tau_\sigma(\vec{r})$ as LSIC($z$) and to  LSIC with $w_{i\sigma}(\vec{r}) = \rho_{i\sigma}(\vec{r})/\rho_\sigma(\vec{r})$ as the weight factors as LSIC($w$), and similarly for OSIC($z$) and OSIC($w$) for OSIC with the corresponding iso-orbital indicator.
The LSIC method is implemented in the development version of the FLOSIC code.\cite{FLOSICcode,FLOSICcodep}

\section{\label{sec:compdetails}Computational Details}

\begin{figure}
\includegraphics[width=0.6\linewidth]{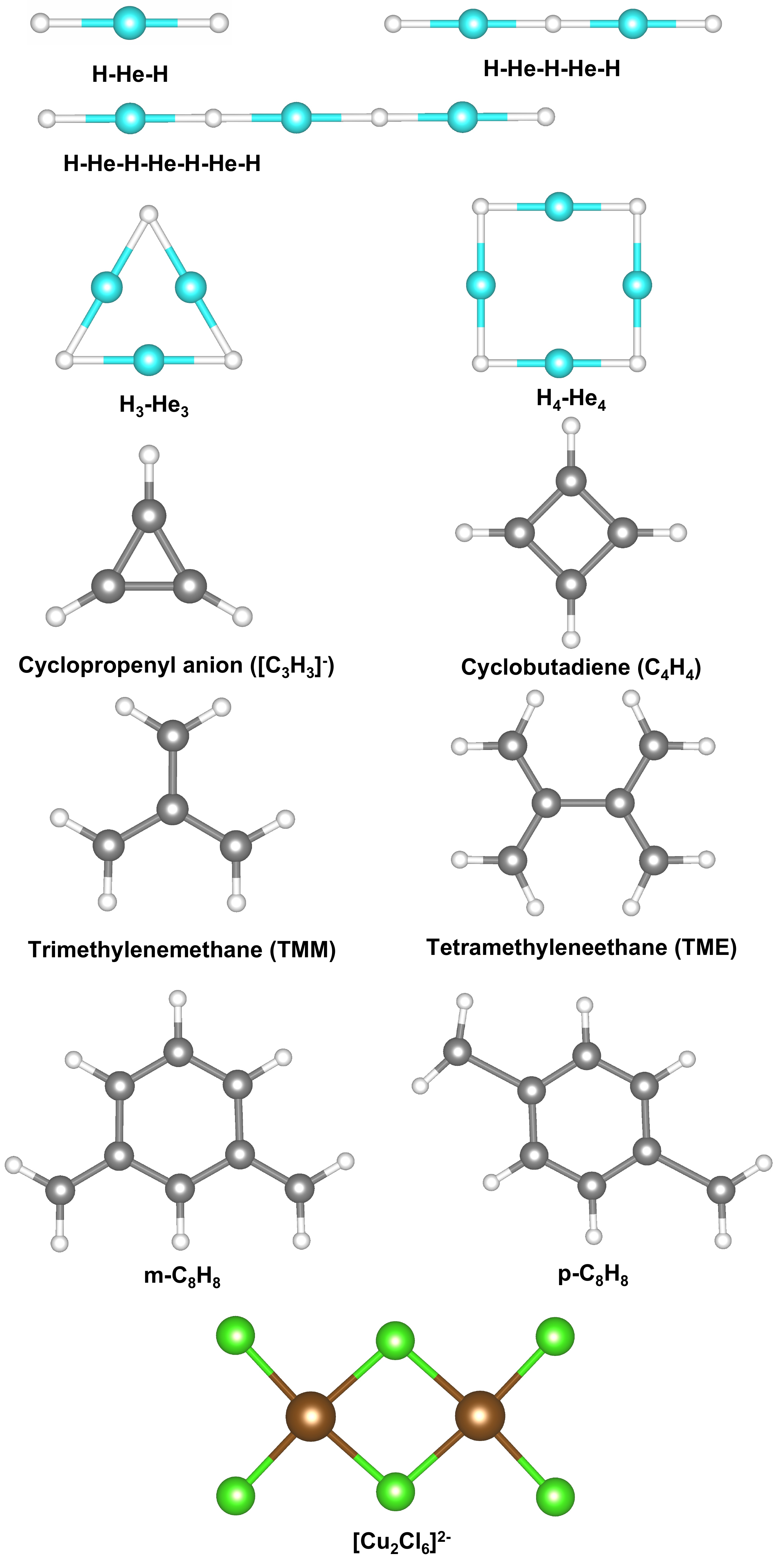}
\caption{\label{fig:geometry}Molecular schemes used in this work for the  evaluation of  magnetic exchange couplings constants with self-interaction corrected methods. 
Hydrogen, helium, carbon, chlorine, copper atoms are color coded as white, cyan, gray, green, and brown, respectively.}
\end{figure}

All calculations in this work were performed with the development version of the FLOSIC code.\cite{FLOSICcode,FLOSICcodep} 
The FLOSIC code is a software based on the UTEP-NRLMOL code and inherits the features of the original 
NRLMOL including the NRLMOL Gaussian-type orbital basis set\cite{PhysRevA.60.2840} and adaptive
integration mesh.\cite{PhysRevB.41.7453} 
This code has the FLO implementation of PZSIC, OSIC, and LSIC. Hereafter, we refer to FLO-PZSIC, FLO-OSIC, and FLO-LSIC as PZSIC, OSIC, and LSIC for brevity.
 To make calculations attainable, in this work we employ a perturbative flavor of LSIC,
 where the PZSIC densities are utilized to evaluate the LSIC energy. 
In case of OSIC method we have found that one shot OSIC results to be practically same as 
fully self-consistent  OSIC results\cite{romero2021local}.

We considered the following representative exchange-correlation functionals:
the local spin density approximation (LSDA) parameterized by Perdew and Wang (PW92)\cite{PhysRevB.45.13244}, the 
 Perdew-Burke-Ernzerhof (PBE)\cite{perdew1996generalized,PhysRevLett.78.1396} generalized gradient 
 approximation (GGA) functional, and the strongly constrained and appropriately normed (SCAN)\cite{sun2015strongly} 
 and the recent r$^2$SCAN meta-GGA\cite{furness2020accurate} functionals. Since SCAN functional is numerically problematic, we used a very dense integration mesh
(roughly 45000 grid points per atom for organic molecules)
 specifically tailored  for  this functional to obtain  accurate energies.\cite{doi:10.1063/1.5120532}
Likewise, we used a mesh with an intermediate grid density between GGA and SCAN for
r$^2$SCAN calculations since meta-GGAs generally require finer meshes than GGAs.
 
 The PZSIC is applied in combination with the LSDA, PBE and SCAN functionals.
For the LSIC calculations, we considered only LSDA  since LSIC combined with GGAs 
or meta-GGAs causes a gauge inconsistency problem\cite{doi:10.1063/5.0010375}.

We first assess FLOSIC and locally-scaled SIC methods for  magnetic exchange coupling constants using 
H$\cdots$He model systems, which were introduced to study magnetic exchange coupling parameters 
with different electronic structure methods, as well as  representative realistic molecular systems, 
including a small transition metal complex.
For all the H$\cdots$He models  reference values from accurate wave function calculations are available.
We included a linear H--He--H molecule of three H--He distances 
$d=1.25, 1.625,$ and $2.00$ \AA, two H--(He--H)$_n$ chains with $n=2$ and $3$, a triangle and a square (see Fig.~\ref{fig:geometry}).
For all these model systems, the 1$s$ hydrogen orbitals are coupled through a super-exchange mechanism mediated by the He orbitals. 
These simple super-exchange  model systems were
used multiple times in the literature to  assess the performance of electronic structure methods for magnetic exchange couplings.
The smallest linear H--He--H has been also used to address the effect of SIE on exchange coupling constants.\cite{doi:10.1063/1.2085171} 
Additionally, we  considered six organic bi-radical molecules:  [C$_3$H$_3$]$^-$, C$_4$H$_4$, trimethylenemethane (TMM), p-C$_8$H$_8$, m-C$_8$H$_8$, and tetramethyleneethane (TME), and chlorocuprate [Cu$_2$Cl$_6$]$^{2-}$ (shown   in Fig.~\ref{fig:geometry}).
The structures of [C$_3$H$_3$]$^-$ and C$_4$H$_4$ were taken from Ref.~\citenum{saito2011symmetry}, 
and the structures of p-C$_8$H$_8$ and m-C$_8$H$_8$ are from Ref.~\citenum{doi:10.1021/acs.jctc.5b00349}. 
The TMM and TME structures  are taken from Ref.~\citenum{joshi2018fermi}. 
The [Cu$_2$Cl$_6$]$^{2-}$ structure was taken from
Ref.~\citenum{doi:10.1021/ic961448x}.  

The NRLMOL Gaussian basis set was used for H/He models  and the organic radical molecules.\cite{PhysRevA.60.2840} 
For the  [Cu$_2$Cl$_6$]$^{2-}$ complex, the Stuttgart relativistic small core effective 
core potential (ECP) consisting of 10 electrons 
was used in combination with its corresponding valence basis set\cite{dolg1987a,martin2001a} for copper, and
Stuttgart relativistic large core ECP was used for chlorine atoms.
The Stuttgart basis set and ECP parameters were obtained from the Basis Set Exchange library.\cite{pritchard2019a}

The optimal FLOs in the FLOSIC calculations are obtained by minimizing the 
SIC energy with respect to the FOD positions.
We used the conjugate gradient algorithm to optimize FOD positions. 
The force tolerance of  $10^{-3} E_h/a_0$ was used for the organic and copper molecules. 
A tighter tolerance of 
$10^{-5} E_h/a_0$ was used for FOD optimization for the H-He complexes.
 A typical relaxed FOD structure has
 an FOD at the nuclear position  corresponding to the $1s$ electrons for atoms other than one-electron systems. For  carbon and oxygen, the valence FODs are found at the vertices of a  tetrahedron centered at  the corresponding nuclei for the hybridized $2s2p$ electrons.
 The valence electrons in the $3s3p3d$ shells in the copper atoms form  symmetrical arrangements around the atomic centers.
The set of optimized FOD used in this work is available on GitHub (https://github.com/FLOSIC/si\textunderscore magnetic\textunderscore exchange\textunderscore coupling).

For the magnetic exchange coupling calculations,  
both $J_{SP}$ and $J_{NP}$ were obtained using Eqs. (\ref{eq:Jsp}) and (\ref{eq:Jnp}), 
where each $E_{HS}$ and $E_{BS}$ were calculated at the 
DFA, PZSIC-DFA, OSIC-DFA, and LSIC-DFA levels of theory.
Since the HS and BS states have different orbital  configurations (i.e. one for parallel and another for anti-parallel spins),
in our FLOSIC calculations
we obtained a distinct set of optimal FODs for each spin configuration: one corresponding to the HS state and the other to the BS state.

\section{\label{sec:results}Results and Discussion}
\begin{table*}
\caption{\label{tab:H2He} Magnetic exchange coupling constants $J_{SP}$ 
(in cm$^{-1}$) for H–He–H with the three H–He distances, $d$. $J_{NP}$ is a half of $J_{SP}$.
Mean absolute percentage deviations (MAPD) with respect to full-CI are also shown.}
\begingroup
\renewcommand{\arraystretch}{1}
\setlength{\tabcolsep}{3.0pt} 
\begin{tabular*}{0.98\textwidth}{@{\extracolsep{\fill}}cccccc}
\toprule
 Method &\multicolumn{3}{c}{$d$ (\AA)}& \multicolumn{2}{c}{MAPD(\%)}\\ \cmidrule(lr){2-4}\cmidrule(lr){5-6}
       & 1.25 & 1.625 &  2.00 & SP & NP\\ \hline
 LSDA  & -12493 & -1494 & -159 & 183 & 41\\
PBE  & -8672 & -916 & -88 &74 & 13\\
SCAN   &-8246 & -948 & -80 &68 &16\\
r$^2$SCAN   &-8304 & -956& -84 &72 &14\\
 \hline
PZSIC-LSDA  & -5503& -632 & -61 &17 &42\\
PZSIC-PBE  & -4894& -541& -51 &1 &50\\
PZSIC-SCAN  & -5526 & -606& -48 &10 &44\\
PZSIC-r$^2$SCAN & -5600 & -633 & -51 & 11 &44\\
 \hline
LSIC($z$)-LSDA (perturbative) & -5734 &-696 & -76 &33 &34\\
LSIC($w$)-LSDA (perturbative) &-4786&-429&-24 &25 &62\\
\hline 
OSIC($z$)-LSDA (perturbative) & -5820 & -709 & -77 &35 &33\\
OSIC($z$)-LSDA (SCF) & -5775 & -683 & -60 &22 &39\\
OSIC($w$)-LSDA (perturbative)  & -5318 &-590 & -54 &9 &45\\
OSIC($w$)-LSDA (SCF)  & -5332 &-606 & -58 &12 &43\\
 \hline
 LSDA@PZSIC-LSDA & -8427 & -1042 & -109 & 94 &9 \\
PBE@PZSIC-PBE &  -6599 & -767  & -76 & 42 & 28\\
SCAN@PZSIC-SCAN & -7240 &-835 & -71 &48 &25\\
r$^2$SCAN@PZSIC-r$^2$SCAN &  -7213 & -845 & -74&50 &24\\
  \hline
Full-CI$^\text{a}$  & -4860 &-544 & -50 & --- & ---\\
\bottomrule
\end{tabular*}

\begin{flushleft}
$^\text{a}$Reference \citenum{doi:10.1021/ic961448x}\\
\end{flushleft}
\endgroup
\end{table*}

\subsection{H--He--H model}

For the linear H--He--H model system, we considered three different H--He bond distances ($d=$1.25, 1.625, and 2.00 \AA) as mentioned in Sec.~\ref{sec:compdetails} with the  LSDA, PBE, and SCAN and r$^2$SCAN meta-GGAs functionals and their SIC counterparts.
In Table \ref{tab:H2He}, we show the calculated $J_{SP}$  
for the H--He--H model with DFAs, PZSIC-DFAs, LSIC, OSIC, DFA@PZSIC-DFA,
where DFA@PZSIC-DFA denotes PZSIC density with a DFA,
and reference full-configuration interaction values.\cite{doi:10.1021/ic961448x}
The mean absolute percentage deviations (MAPD) for each method are also included in Table \ref{tab:H2He}. 
It should be mentioned 
that for this system, the SP method to map exchange couplings is physically more meaningful than 
the NP since it takes into account the broken spin symmetry nature of the low-spin state. 
By contrast, the NP approach assumes that the broken symmetry state provides the correct energy of the low spin state. 
Since in this case the spin magnetization is very localized at the H centers, 
the SP method is expected to yield the correct mapping. The NP values are half of the SP values for this system and therefore are  not shown.
Typically, DFA exchange couplings using SP formulation are overestimated, and their $J_{NP}$ values are therefore closer to experimental values. \cite{phillips2011role}

For all the four DFA functionals employed here, we find that the $J_{SP}$ values are  larger by a factor of $2-3$ compared to the reference values. 
When SIE are removed using PZSIC, the predicted antiferromagnetic coupling strengths decrease, 
resulting in reduction in the deviations in all three cases.
$J_{SP}$'s obtained with PZSIC-PBE show the smallest MAPD, with results 
comparable to the full-CI reference values.

\begin{figure}
\includegraphics[width=1\linewidth]{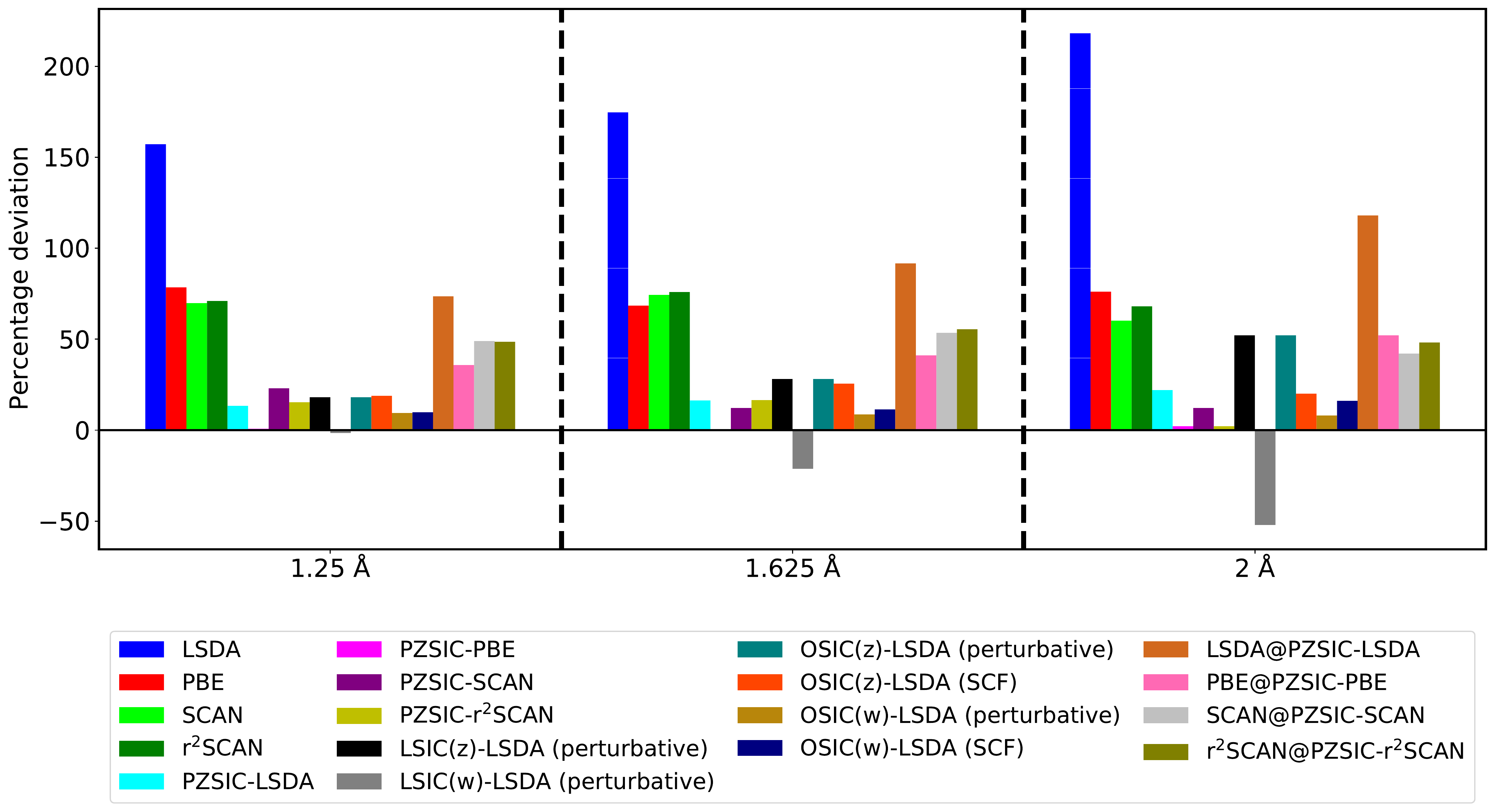}
\caption{\label{fig:JSP_H-He_H} Percentage deviations of $|J_{SP}|$ 
for H--He--H with respect to the CI values.}
\end{figure}

A graphical comparison of the percentage deviations of $J_{SP}$'s is shown in Fig. \ref{fig:JSP_H-He_H}.
PZSIC with the four DFAs shows fairly good performance with relatively small deviations (MAPD, $1-17\%$).  
The reduction of $J_{SP}$ with PZSIC is partly due to an improved  electron density. To examine the effect of PZSIC density, we calculated the $J_{SP}$ parameters using the PZSIC density with the DFAs (herein DFA@PZSIC-DFA). 
The calculated DFA@PZSIC-DFA values are also shown in Table~\ref{tab:H2He}.  All the DFA@PZSIC-DFA values are intermediate between respective DFA and PZSIC-DFA values. Moreover, the comparison of DFA errors with DFA@PZSIC-DFA errors shows that the
density correction accounts for $17-49\%$ of the errors in DFA-only calculations going from r$^2$SCAN to LSDA.  Thus the density correction plays a significant part in improving the 
results particularly for LSDA and GGA functionals which belong to the lower rungs of Jacob ladder of density functionals.\cite{doi:10.1063/1.1390175} 
The rest of the correction in PZSIC arises from the correction to the energy functional.
Since the H--He--H system consists of mostly single electron regions where PZSIC is exact, 
PZSIC performs well for $J_{SP}$ for this set of systems. On the other hand, the LSIC($z$), LSIC($w$) and OSIC($z$) methods surprisingly perform much worse compared to PZSIC. LSIC methods are applied perturbatively on PZSIC density and therefore the higher errors in these results stem from the functional alone. LSIC mimics PZSIC in the one-electron regions but reduces the corrections in the many-electron regions. 
As LSIC reduces the SIC energy correction with respect to PZSIC,
the LSIC total energies are higher than the PZSIC total energies.
This energy shift is larger on the HS states than the BS states, and hence LSIC shows larger values of the coupling constant than PZSIC. The best performance for this system comes from the OSIC($w$) method. The $w$ factor identifies the weak interaction regions better compared to the $z$ factor which may have contributed to the better performance of the OSIC($w$).

In the next sections, we investigated systems that involve many-electron regions where PZSIC may show overcorrections.

\subsection{H$\cdots$He multicenter models}

Next, we assessed $J$ couplings with SIE-free DFAs on a set of multi-center H$\cdots$He model systems the structures of which are shown in Fig. \ref{fig:geometry}.
These complexes are  extensions of the two-center H--He--H for which 
accurate wave function as well as hybrid functional calculations are available.\cite{ruiz2003calculation} 
Here, we studied the four multi-center complexes: H$_3$He$_2$, H$_4$He$_3$, H$_3$He$_3$, and H$_4$He$_4$. The structures of these complexes are taken from Ref. \citenum{ruiz2003calculation}.
The bond distances between H$\cdots$He were fixed to 1.625\AA{} in all four cases.  
For the evaluation of $J_{SP}$ couplings, Eq.~(\ref{eq:HDVV}) can be easily extended to multicenter spin cases. \cite{ruiz2003calculation} The equations used to determine the coupling constants of the multicenter complexes can be found in supplementary information.

Our calculated $J_{ij}$ for the nearest and the second nearest neighbor interactions with DFAs, PZSIC-DFAs, and LSIC are presented  and compared  against the available CASPT2 values in Table \ref{tab:multicenter}.
All four DFAs overestimate  
$J_{SP}$ with LSDA having the largest errors. 
The mean absolute percentage deviations for PBE, SCAN, and r$^2$SCAN are comparable.
The first nearest neighbor couplings ($J_{SP}^{12}$ and $J_{SP}^{23}$) tend to show smaller absolute percentage deviations than the second nearest neighbor couplings ($J_{SP}^{13}$) because the relative uncertainties are small. 
Removing SIE with PZSIC systematically reduces these deviations with MAPDs of 
27, 24, and 38\% for PZSIC-LSDA, -PBE, and -SCAN, respectively. 
On the other hand, LSIC($z$) tends to increase the  coupling strengths for the nearest neighbor pairs and thus results in  overestimation, consistent  with what we observed for the linear H--He--H.
For the second nearest neighbor pairs, which are weakly coupled, LSIC($z$) tends to improve the coupling strengths over PZSIC. On the other hand, LSIC($w$) tends to underestimate the coupling in general and even leads to ferromagnetic coupling for the next nearest neighbors.
Interestingly, when MAPDs are compared, the density based OSIC($w$) method shows the smallest percentage deviation of 14\% among the SIC calculations. No systematic improvement is seen for the OSIC($z$) method. Overall, PZSIC-PBE and OSIC($w$) performs well for this set of toy systems.

\begin{table}
\caption{\label{tab:multicenter} Magnetic exchange coupling constants $J_{SP}$ (in cm$^{-1}$) for 
the H$\cdots$He 
multicenter complexes. 
The bond length H$\cdots$He is fixed to 1.625 \AA{} in all cases. Structures are shown in Fig.~1.}
\fontsize{8}{8}\selectfont
\begingroup
\setlength{\tabcolsep}{1.9pt} 
\begin{tabular*}{0.98\textwidth}{@{\extracolsep{\fill}}ccccccccccccccccc}
\toprule
 System  &&\multicolumn{4}{c}{DFA}   &\multicolumn{3}{c}{PZSIC}&LSIC($z$)&LSIC($w$)&\multicolumn{2}{c}{OSIC($z$)}&\multicolumn{2}{c}{OSIC($w$)}& Ref.$^\text{a}$ \\ \cmidrule(lr){3-6} \cmidrule(lr){7-9} \cmidrule(lr){12-13} \cmidrule(lr){14-15}
&& LSDA & PBE & SCAN & r$^2$SCAN & LSDA & PBE & SCAN & LSDA$^\text{b}$ &LSDA$^\text{b}$&LSDA$^\text{b}$& LSDA& LSDA$^\text{b}$& LSDA& \\ 
\midrule
\multirow{2}{*}{H$_3$He$_2$}  &$J_{12}$  &-1673  &-1045 &-994 &-1019& -682 &-600 &-670& -744 &-467&-763&-735&-633 & -647&-586  \\ 
   & $J_{13}$  &-41.3  &-6.7  &-8.4 & -8.9& -3.0 &-1.8 &-2.0&-3.2 &-1.4&-2.8&-4.8&-2.5&-2.9 & -4.0 
  \\ 
\midrule
\multirow{3}{*}{H$_4$He$_3$} &$J_{12}$  &-1687  &-1049 &-1000& -1025& -684 &-602 &-662&-747&-480&-766&-740&-636&-650&-587   \\ 
      &  $J_{23}$ &-1830  &-1188  &-1036&-1060 & -733 &-660 &-702&-786&-494&-812&-781&-681&-696& -629 \\ 
      &  $J_{13}$ &-44.5  &-6.9  &-7.3&-7.8 & -4.7 &-2.3 &-6.6&-3.0&10.0&-3.4&32.6&-3.7&-3.9 &-3.4 \\ 
\midrule
H$_3$He$_3$ &$J_{12}$  &-442  &-314 &-305&-311 &-230 &-221 &-228&-263 &-193&-274&-239&-231&-235 &-197   \\       
\midrule
\multirow{2}{*}{H$_4$He$_4$}   &$J_{12}$  &-1425  &-917 &-838& -849&-611 &-540 &-576& -606&-412&-651&-619&-558&-569 &-516   \\ 
    &  $J_{13}$& -22.3 & 0.5  &-1.5 &-1.5 & -2.5 &-1.6 &-0.5
&-7.2&6.0&-4.3&30.7&-7.5&-9.2&-8.8  \\ 
\midrule
\multirow{2}{*}{MAPD} & SP& 395 & 82& 79 & 84 & 27 & 24 & 38	& 23 & 89 & 30 & 206 &14 &14 &--\\
& NP &147 &22 &24 &25 &48 &59 &49	&45 &94 &45 &136 &50 &47 &--\\
\midrule
\multirow{2}{*}{MAD} & SP& 579 &252	&209 &221 &54 &15 &42 &79 &62 &95 &84 &28 &36 &--\\
& NP &131	&34	&55	&49	&132 &152	&138 &119 &189 &112 &125 &144 &141 &--\\
\bottomrule
\end{tabular*}
\begin{flushleft}
$^\text{a}$CASPT2 values from Ref.~\citenum{ruiz2003calculation}
\\
$^\text{b}$Perturbative calculation using PZSIC-LSDA densities. 
\end{flushleft}
\endgroup
\end{table}

\subsection{Organic radical molecules}

In this section, we assess the performance of SIC approximations for $J$ couplings on a set of six  organic radical molecules,  
 [C$_3$H$_3$]$^{-}$, C$_4$H$_4$, trimethylenemethane (TMM), p-C$_8$H$_8$, m-C$_8$H$_8$, and tetramethyleneethane (TME) (Fig.~\ref{fig:geometry}).
This set is the same set used in 
the work of Joshi \textit{et al.} where FLOSIC-LSDA was used to investigate the effects of SIE on exchange coupling constants.\cite{joshi2018fermi}
We used the same methods as in the previous Section \ref{sec:theory}, with the exchange couplings in this case  compared against available multireference configuration interaction or multi-reference perturbation theories such as CASSCF.\cite{saito2011symmetry,cramer1996trimethylenemethane,rodriguez2000controversial,zhang2000effect,reta2014triplet}
The results are summarized in Table \ref{tab:radmols}.
The SP approach assumes that the energies of the electronic single-reference states can be mapped to the energies of broken-symmetry spin states, which is  a physically acceptable assumption for the H$\cdots$He complexes with localized electrons. This is not the case 
for the organic complexes, since the spin density is mostly delocalized. Hence, for these organic radical molecules, the NP mapping, which assumes that the electronic single-reference states provide a reasonable energy for the actual multi-reference states, is more physically acceptable. 
In  Table \ref{tab:radmols} we show $J_{SP}$ values ($J_{NP}$ values are half as large as the $J_{SP}$ values). The MAPDs of both approaches are shown.

\begin{figure}
\includegraphics[width=\columnwidth]{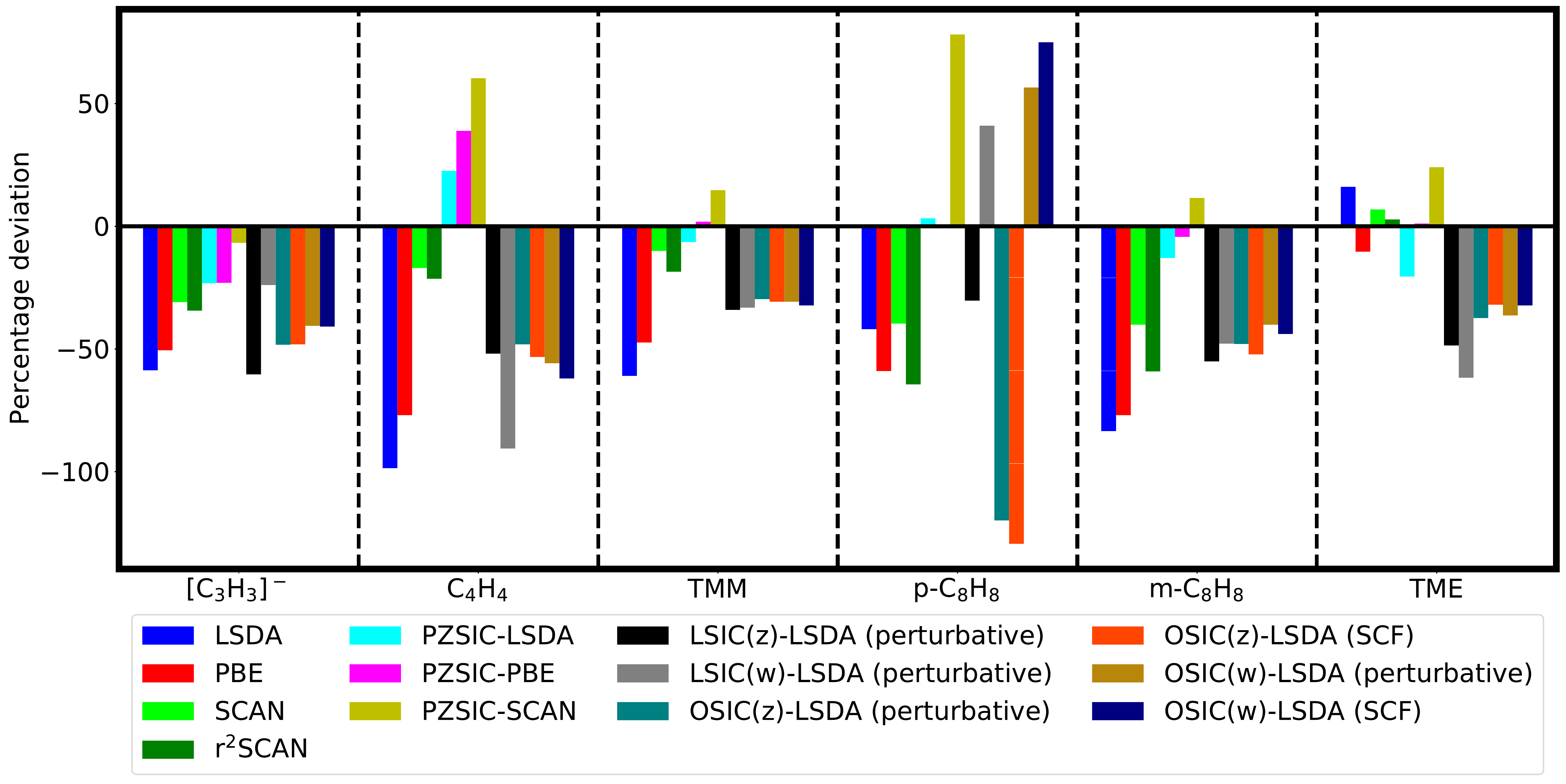}
\caption{\label{fig:np-organic} Percentage deviation of $|J_{NP}|$
for the set of six organic systems with respect to multi-configuration calculations.}
\end{figure}

Similar to the results of the H$\cdots$He systems, the DFA errors for the organic complexes too 
are large with MAPDs ranging from 60\% for LSDA to 24\% for SCAN.   PZSIC improves the results, but the improvement is not consistent for all the DFAs employed here. While significant improvements are seen for LDA and PBE with PZSIC resulting in MAPDs of 14 and 12\%,  SCAN performance deteriorates to MAPD of 33\%. With PZSIC-DFA, the strength of the interaction (either ferro or antiferromagnetic) is enhanced with respect to the parent DFAs couplings  except for TME with PZSIC-LSDA.
It is interesting to observe that the scaled SIC methods assessed here affect exchange couplings in a non-systematic way when compared to non-scaled PZSIC. The signs of the coupling constant are consistent with the reference values with DFAs, PZSIC-DFAs, LSIC($z$), LSIC($w$), and OSIC($w$) but not with OSCI($z$). For p-C$_8$H$_8$, OSIC($z$) favors ferromagnetic coupling while the reference results favor antiferromagnetic coupling. The coupling strengths are generally enhanced with LSIC($z$), LSIC($w$) and OSIC($w$) for all molecules except TME for which all the scaled SIC methods show lower strength compared to LSDA. 
Fig.~\ref{fig:np-organic} shows the percentage deviation between the computed
$|J_{NP}|$ with DFA, PZSIC-DFA, LSIC, and OSIC   and the corresponding reference 
values for each molecule. 
We also compared the coupling constants calculated using the DFA@PZSIC-DFA approach
with the reference data  in Table~\ref{tab:radmols_atcalculation}. These results 
show that the SIC to the density correction has little influence on the
results with LSDA and PBE. The use of the SIC density in these functionals provide 
marginal improvement in MAPD of  1 and 3\%, respectively. It is clear 
however that using SIC  density in more accurate functionals such as 
SCAN and r$^2$SCAN results in appreciable reduction in MAPD by 9 and 16\%.

\begin{table}
\caption{\label{tab:radmols} 
Calculated magnetic exchange coupling constants $J_{SP}$ (in cm$^{-1}$) where $J_{NP}$ is half the $J_{SP}$ for a set of organic radical molecules.}
\fontsize{8}{8}\selectfont 
\setlength{\tabcolsep}{1.9pt} 
\begin{tabular*}{0.98\textwidth}{@{\extracolsep{\fill}}cccccccccccccccc}
\toprule
 \multicolumn{2}{c}{System}  &\multicolumn{4}{c}{DFA}   &\multicolumn{3}{c}{PZSIC}&LSIC($z$)&LSIC($w$) &\multicolumn{2}{c}{OSIC($z$)}&\multicolumn{2}{c}{OSIC($w$)}& Ref. \\ \cmidrule(lr){3-6} \cmidrule(lr){7-9} \cmidrule(lr){12-13} \cmidrule(lr){14-15}
&& LSDA & PBE & SCAN & r$^2$SCAN & LSDA & PBE & SCAN & LSDA$^\text{a}$ &LSDA$^\text{a}$ & LSDA$^\text{a}$ & LSDA &LSDA$^\text{a}$ & LSDA &\\ 
\midrule
\multicolumn{2}{c}{
[C$_3$H$_3$]$^-$}  
&3759  &4506 &6286 &5974 & 6990 &7004 &8490 &3610 & 6923 &4712   &4720   &5406 &5382 & 4547$^\text{b}$\\ 
\midrule
\multicolumn{2}{c}{C$_4$H$_4$}  
&-88  &-1306 &-4704 &-4546 &-6940 &-7868 &-9084 &-2734 &-538 &-2945   &-2653   &-2508 &-2154 & -2833$^\text{b}$\\ 
\midrule
\multicolumn{2}{c}{TMM}  
 &4798  &6448 &11022 &9982 &11487 &12470 &14052 &8078 &8196 &8614 &8488   &8484 &8294 &6121$^\text{c}$\\ 
\midrule
\multicolumn{2}{c}{p-C$_8$H$_8$}   
 &-2710  &-1916 &-2816& -1666&-4816 &-4662 &-8308 &-3258 &-6573 &922 &1367   &-7302 &-8162 &-2333$^\text{d}$   \\ 
\midrule
\multicolumn{2}{c}{m-C$_8$H$_8$}   
  &2276 &3146 &8176 &5580 &11904 &13078 &15212 &6136 &7134 &7102   &6536   &8194 &7668 &6824$^\text{e}$  \\ 
\midrule
\multicolumn{2}{c}{TME}   
&-2838 &-2198 &-2614 &-2516 &-1948 &-2486 &-3036 &-1260 &-938 &-1534  &-1668   &-1562 &-1660 &-1224$^\text{f}$  \\ 
\midrule
\multirow{2}{*}{MAPD} & SP &58	&35	&56	&51	&87	&105	&161	&18 &63&36&41	&55	&63&\\
&NP &60	&53	&24	&33	&14	&12	&33	&47	&50&55&58   & 43	&48&\\
\bottomrule
\end{tabular*}

\begin{flushleft}
$^\text{a}$Evaluated using the PZSIC-LSDA density\\
$^\text{b}$CASSCF-MkCCSD from reference \citenum{saito2011symmetry}\\
$^\text{c}$CASPT2N from reference \citenum{cramer1996trimethylenemethane}\\
$^\text{d}$AM1-CI from reference \citenum{zhang2000effect}\\
$^\text{e}$CASSCF from reference \citenum{reta2014triplet}\\
$^\text{f}$DDCI from reference \citenum{rodriguez2000controversial}\\
\end{flushleft}
\end{table}

\begin{table}
\caption{\label{tab:radmols_atcalculation}
Calculated magnetic exchange coupling constants $J_{SP}$ in cm$^{-1}$ using DFA@PZSIC-DFA for the set of organic radical molecules.}
\fontsize{9}{9}\selectfont  
\begin{tabular*}{0.98\textwidth}{@{\extracolsep{\fill}}cccccc}
\toprule
 \multicolumn{2}{c}{System}  & LSDA@PZSIC-LSDA & PBE@PZSIC-PBE & SCAN@PZSIC-SCAN&r$^2$SCAN@PZSIC-LSDA\\
\midrule
\multicolumn{2}{c}{[C$_3$H$_3$]$^-$}  
&4001& 5024&	6466&	5712\\ 
\midrule
\multicolumn{2}{c}{C$_4$H$_4$}  
&-713&-2340&	-5408&	-5342\\ 
\midrule
\multicolumn{2}{c}{TMM}  
&5902&7638&	10866&	9748\\ 
\midrule
\multicolumn{2}{c}{p-C$_8$H$_8$}   
&-2557&-1308&	-3464&	-4506 \\ 
\midrule
\multicolumn{2}{c}{m-C$_8$H$_8$}   
&4896 &6596&	15031&	9274  \\ 
\midrule
\multicolumn{2}{c}{TME}   
&-1305 &1494	&-2686&	-2396 \\ 
\midrule
\multirow{2}{*}{MAPD} &SP &22 &20&83&66\\
&NP &59 &50&15&17\\
\bottomrule
\end{tabular*}
\end{table}

\subsection{Hexa-Chlorocuprate [Cu$_2$Cl$_6$]$^{2-}$}
\begin{figure}
    \centering
    \includegraphics[width=0.8\linewidth]{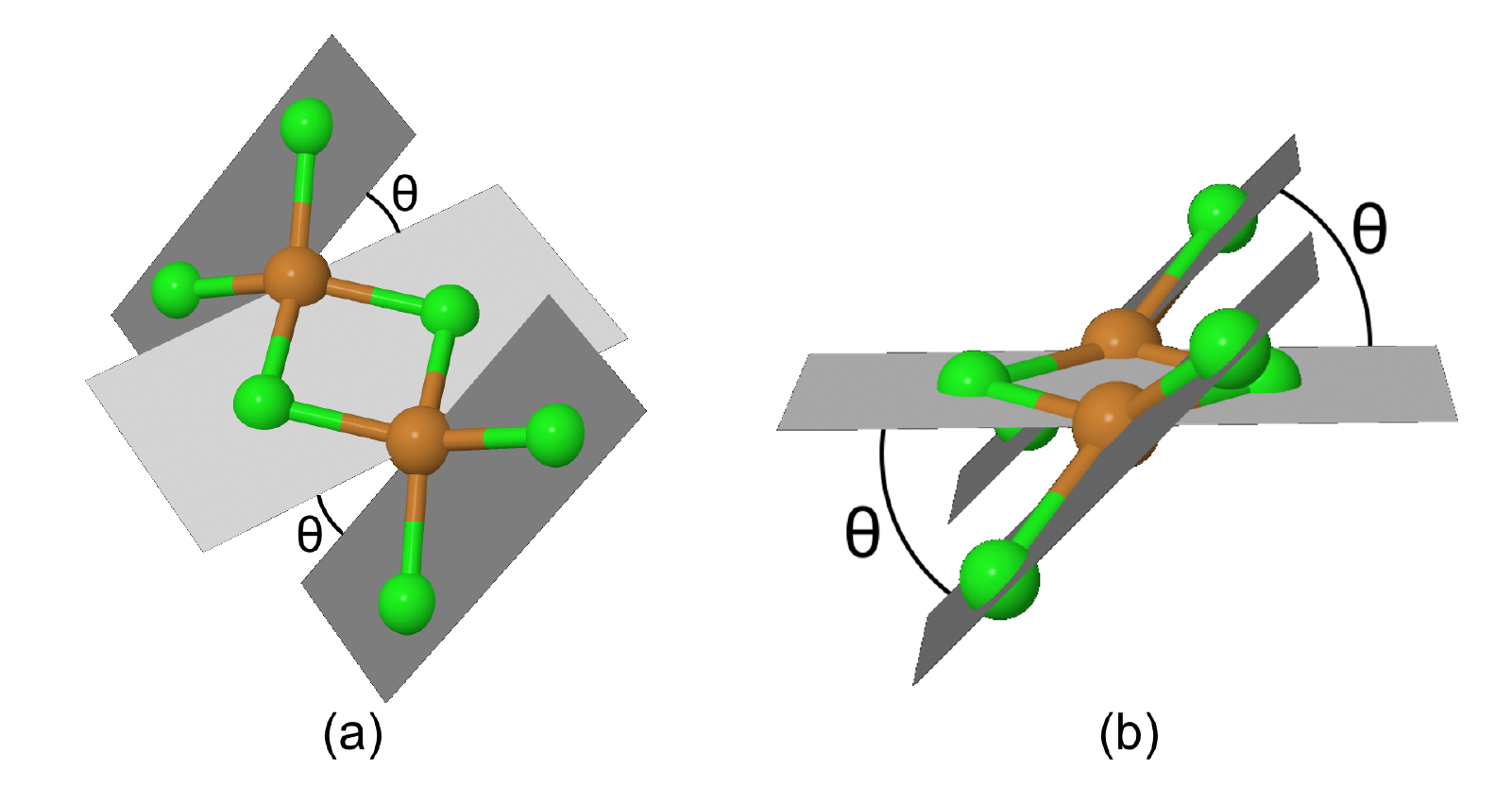}
    \caption{The structure of chlorocuprate [Cu$_2$Cl$_6$]$^{2-}$ where the outer CuCl$_2$ planes are twisted by angle $\theta$ with respect to the plane made by the inner Cu$_2$Cl$_2$ square: (a) top down view and (b) side view.
    }
    \label{fig:cu2cl6visual}
\end{figure}

The hexa-chlorocuprate  complex has two Cu(II) bridged with chlorine ligands, as shown in Fig.~\ref{fig:cu2cl6visual}, 
with each copper atom in a $3d^9$ electron configuration that effectively acts as a spin $S=1/2$ site.
Experimentally, the chlorocuprate [Cu$_2$Cl$_6$]$^{2-}$ complex is found to be weakly antiferromagnetic.\cite{willett1985magneto}
The  magnetic susceptibility analysis of this complex shows that its magnetic character
 changes when
outer non-bridging chlorine atoms are twisted around the Cu-Cu axis by an angle $\theta$ (see Fig. \ref{fig:cu2cl6visual}).\cite{doi:10.1021/ic00199a014}
As  $\theta$ varies from $0\degree$ to $45\degree$, the coupling  changes from being 
weakly antiferromagentic to ferromagnetic. 
This change in the nature of the magnetic interaction with the angle $\theta$ makes the evaluation of the exchange coupling particularly challenging for DFT methods.
Relative to the other organic complexes and the H-He model systems studied in this work, the $d$ electrons in this 
complex are expected to be largely affected by the SIEs, 
making the hexa-chlorocuprate an interesting case to  study the effect of SIE removal  on magnetic  exchange couplings. 
To this end, we use two 
structures taken from the literature.\cite{doi:10.1063/1.1430740,doi:10.1021/ic961448x,doi:10.1021/ic961448x} 
The first structure is the planar [Cu$_2$Cl$_6$]$^{2-}$ 
($\theta=0\degree$)\cite{doi:10.1063/1.1430740,doi:10.1021/ic961448x}
for which experiments  show that the  $J$ values  are between 0 to $-$40 cm$^{-1}$, indicating that the Cu(II) ions are weakly antiferromagnetically coupled.\cite{willett1985magneto} 
This antiferromagnetic coupling constant is close to the value determined with PBE functional for Cu$_3$ complex in which case the Cu atoms are further apart \cite{PhysRevB.82.155446}. 
The second structure for [Cu$_2$Cl$_6$]$^{2-}$ ($\theta=45\degree$) was taken from from  
Bencini \textit{et al.},\cite{doi:10.1021/ic961448x} where the experimental coupling was found between 80 and 
90~cm$^{-1}$. 
This qualitative change in the magnetic interaction with the twisting angle has also confirmed by
{\it ab initio} methods (see Table~\ref{tab:Cu2Cl6}).\cite{doi:10.1021/ic00199a014}
We note that a slightly different planar structure was used in a previous FLOSIC-LSDA study.\cite{doi:10.1063/1.1430740}

The results of our calculations are summarized in Table \ref{tab:Cu2Cl6}. For the $\theta=0\degree$ configuration, most methods give an antiferromagnetic interaction, with the exception of the OSIC with scaling factor $w$. In some of the SIC cases, the interaction strength is exaggerated, but never at the level of LSDA or PBE.   
It is interesting to point out that using the PZSIC densities with their parent DFA largely reduces the strength 
of the coupling for LSDA and PBE and only minimally affects the SCAN results.  
From the scaled methods, the orbital scaled and local scaled SIC LSDA yield similar small antiferromagnetic couplings, in line with 
the experimental and ab initio results.  Turning to the  $\theta=45\degree$ conformation, we found that while most methods predict a ferromagnetic coupling (the exceptions are PZSIC-LSDA and PZSIC-PBE), in some cases the couplings are largely overestimated. Here again the orbital scaled and local scaled SIC LSDA give $J$ couplings in the range of the reference data.

\begin{figure}
    \centering
    \includegraphics[width=0.8\linewidth]{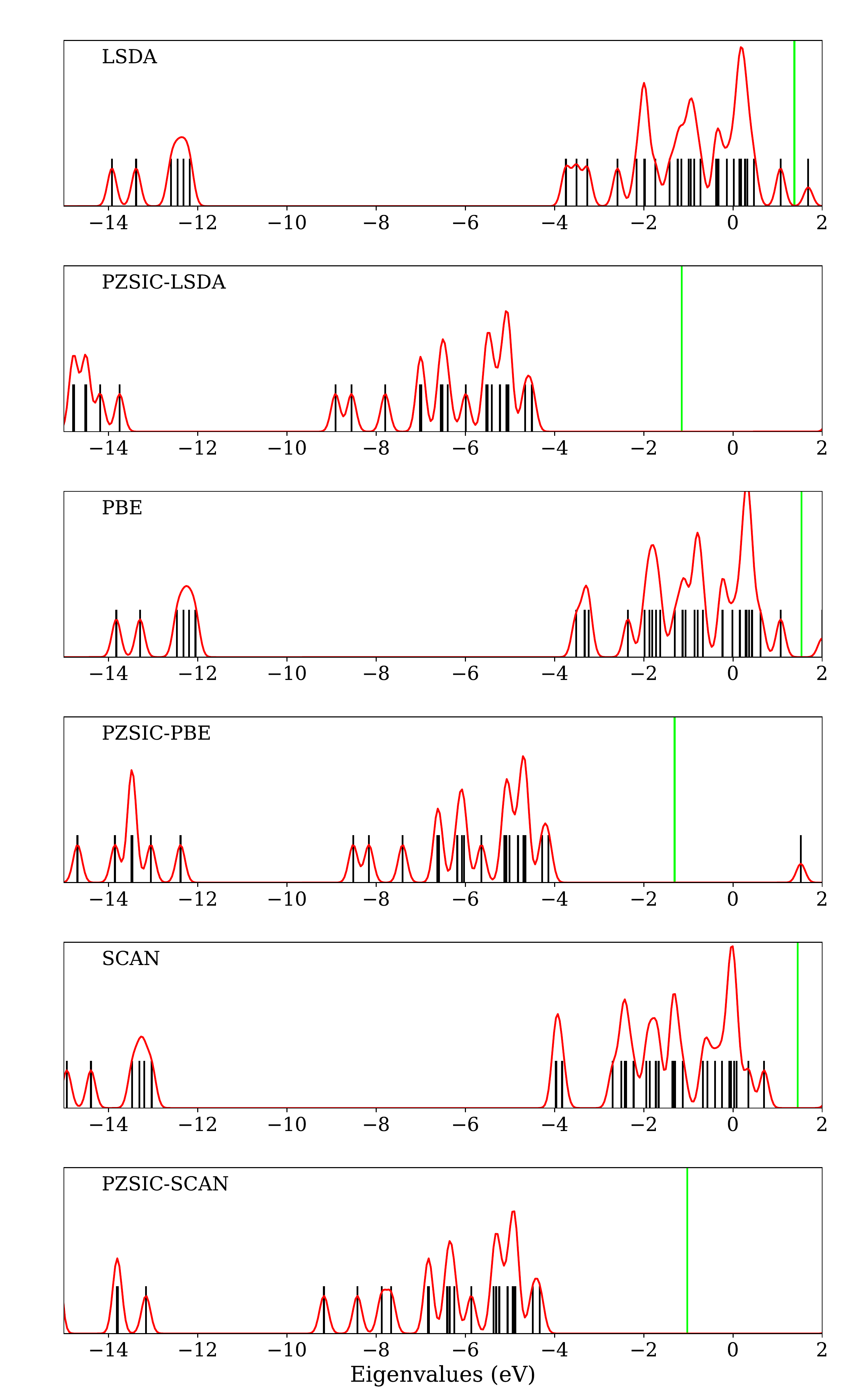}
    \caption{The eigenvalue spectra (in eV) of chlorocuprate [Cu$_2$Cl$_6$]$^{2-}$ in BS states. The green lines indicate the energy levels in between HOMO and LUMO eigenvalues.
    }
    \label{fig:eigenvaluespectra}
\end{figure}

Although for $\theta=0\degree$ configuration SCAN and r$^2$SCAN predict coupling 
constants within the experimental range of values, these results are not completely reliable as 
for these self-interaction uncorrected functionals, the highest-occupied valence electron(s) 
are not bound as evidenced from the 
eigenvalue spectra shown in Fig. \ref{fig:eigenvaluespectra}. It can be seen that removal of SIE 
results in the binding of the valence electrons and thereby 
providing significantly better description of the electron density compared to
the bare 
DFAs. The SIC fixes the excessive electron delocalization
in the uncorrected DFAs as evidenced in the localization of spin-densities on the
Cu atoms as summarized in Table \ref{tab:Cu2Cl6-population}. These spin-populations were obtained
from the Mulliken and L\"owdin population analysis and by integrating the spin 
densities in atomic spheres of van der Waal radii placed on Cu cites.\cite{https://doi.org/10.1002/pssb.200541490,PhysRevB.59.R693} 
Interestingly, the spin populations also systematically increase
in the corrected functionals from LSDA to 
PBE-GGA to SCAN meta-GGA.

Among all the variations of scaled SIC methods analyzed, only LSIC($z$)-LSDA, OSIC($z$)-LSDA are able to capture the correct nature and strength of the magnetic interaction for  $\theta=0\degree$ and $\theta=45\degree$. This is encouraging since these methods
(especially) LSIC also 
perform very well for other properties such as molecular dissociation energies and electric polarizabilities.\cite{doi:10.1063/5.0010375,waterpolarizability,akter2021well}
On the other hand, within the non-scaled SIC methods, SCAN@PZSIC-SCAN and all the r$^2$SCAN@PZSIC-DFA methods also 
successfully reproduce the magnetic interactions. 
This is an  indication that the SIC densities are in general of better quality than their non-corrected counterparts, and these densities used  
with a ``higher-rung'' energy density functional are a good option for the evaluation of magnetic exchange couplings.

\begin{table*}[ht]
\caption{\label{tab:Cu2Cl6}
Magnetic exchange coupling constant $J_{SP}$ (where $J_{NP}$ is a half of $J_{SP}$) in cm$^{-1}$  of the {[Cu$_2$Cl$_6$]$^{-2}$} molecule.
}
\fontsize{12}{12}\selectfont 
\begin{tabular*}{0.98\textwidth}{@{\extracolsep{\fill}}ccc}
\toprule
 Method & $\theta=0\degree$ &  $\theta=45\degree$  \\
 \hline
 LSDA   & -354 & 221   \\
 PBE    & -234 & 150   \\
 SCAN   & -3   & 159   \\
 r$^2$SCAN & -42 & 177  \\
 \hline
 PZSIC-LSDA & -78 & -12  \\
 PZSIC-PBE  & -94 & -24  \\
 PZSIC-SCAN & -99 & 228  \\
 \hline
 LSIC($z$)-LSDA$^\text{a}$ & -3  & 102  \\
 LSIC($w$)-LSDA$^\text{a}$ & 104 & 179  \\
 \hline
 OSIC($z$)-LSDA$^\text{a}$ & -15 & 107   \\
 OSIC($z$)-LSDA(SCF) & -132 & 71  \\ 
 OSIC($w$)-LSDA$^\text{a}$ &  145 & 202  \\
 OSIC($w$)-LSDA(SCF) &  50 & 102  \\
\hline
LSDA@PZSIC-LSDA & -131 & 59  \\
PBE@PZSIC-PBE   & -138 & 80  \\
SCAN@PZSIC-SCAN & -25  & 87  \\
\hline
r$^2$SCAN@PZSIC-LSDA & -49 & 76   \\
r$^2$SCAN@PZSIC-PBE & -87  & 107  \\
r$^2$SCAN@PZSIC-SCAN & -40 & 88   \\
\hline
Ab initio, LOC$^\text{b}$ & -36 & 58  \\
Ab initio, DELOC$^\text{b}$ & -6 & 47 \\
 \hline       
Expt.  &   0 to -40$^\text{c}$ & 80 to 90$^\text{d}$ \\
\bottomrule
\end{tabular*}

\begin{flushleft}
$^\text{a}$Perturbative calculation on PZSIC-LSDA density\\
$^\text{b}$References \citenum{MIRALLES1992555,CASTELL1994377}\\
$^\text{c}$Reference \citenum{willett1985magneto}\\
$^\text{d}$Reference \citenum{doi:10.1021/ic00199a014}\\
\end{flushleft}
\end{table*}

\begin{table*}[ht]
\caption{\label{tab:Cu2Cl6-population} 
Spin density analysis of the {[Cu$_2$Cl$_6$]$^{-2}$} molecule. $\int ( \rho_\uparrow(\vec{r}) - \rho_\downarrow(\vec{r}) ) d\vec{r}$ of BS states at the one of the copper atom is shown. Similar results are found for HS states.
}
\begin{tabular*}{0.98\textwidth}{@{\extracolsep{\fill}}ccccccc}
\toprule
 Method & \multicolumn{3}{c}{$\theta=0\degree$} &  \multicolumn{3}{c}{$\theta=45\degree$}  \\\cmidrule(lr){2-4} \cmidrule(lr){5-7}
 & Mulliken & L\"owdin & Atom sphere  & Mulliken & L\"owdin & Atom sphere\\
 \hline
LSDA	&0.41	&0.48	&0.44	&0.45	&0.51	&0.48\\
PBE	    &0.46	&0.53	&0.49	&0.49	&0.55	&0.52\\
SCAN	&0.54	&0.61	&0.58	&0.57	&0.63	&0.61\\
r$^2$SCAN	&0.52	&0.59	&0.57	&0.55	&0.61	&0.59\\
\hline
PZSIC-LSDA	&0.73	&0.79	&0.78	&0.77	&0.81	&0.80\\
PZSIC-PBE	&0.78	&0.82	&0.81	&0.81	&0.85	&0.84\\
PZSIC-SCAN	&0.76	&0.80	&0.78	&0.80	&0.83	&0.82\\
\bottomrule
\end{tabular*}
\end{table*}

\section{\label{sec:conclusions}Conclusion}
We have studied the magnetic exchange coupling parameters for sets of molecules: 
H--He models, six organic radical molecules, and [Cu$_2$Cl$_6$]$^{2-}$ using 
DFAs with and without SIC scheme of PZSIC.
In addition, we assessed the impact of orbital-scaled SIC methods, such as OSIC and LSIC applied to LSDA to study the performance of these SIC methods on calculating the said property.
For the H--He--H system using the SP method, application of PZSIC improves the resulting exchange couplings especially for PZSIC-PBE. With LSIC-LSDA, performance is slightly worsened compared to PZSIC-LSDA.
For the H--He multi-nuclear systems, we find that removing SIE decreases coupling strengths. PZSIC shows good performance for the three functionals. LSIC($z$)-LSDA shows slightly larger discrepancies than PZSIC-LSDA for the nearest-neighbor couplings but shows better agreement for the second nearest-neighbor couplings with the reference, resulting in a relatively small MAPD for this set. 
The smallest MAPD is seen in the density based OSIC($w$)-LSDA 
with SP and PZSIC-PBE with NP. 
For the set of organic molecules, $J_{SP}$ of uncorrected DFAs shows a fair performance while removing SIE overestimates the values notably.
However, $J_{NP}$ from
PZSIC-DFAs coincidentally provides good agreement with the references where PZSIC-PBE and PZSIC-LSDA show two of the smallest percentage deviations.
Although we find that the $J_{NP}$ approach with LSIC and OSIC performs rather poorly,
the $J_{SP}$ approach combined with LSIC($z$)-LSDA provides the best estimate among all SP approaches (MAPD, $18\%$) in terms of percentage deviations (MAPD of the rest of the methods ranging $35-161 \%$)
where the triplet-singlet energy gaps of PZSIC are reduced in all six systems.
Finally, for the chlorocuprate [Cu$_2$Cl$_6$]$^{2-}$, 
LSIC($z$)-LSDA, SCAN, and r$^2$SCAN produce very weak antiferromagnetic couplings for $\theta=0\degree$ and ferromagnetic coupling for $\theta=45\degree$ showing close agreement with experiments. 
Our DFA@PZSIC-DFA results are, in general, better than those of uncorrected DFAs but not as good as PZSIC-DFAs.
Thus it seems that using an SIC density only is not sufficient and that an SIC energy correction is also needed for this property.
Among the uncorrected DFAs, SCAN outperforms than LSDA and PBE in all cases.
Also, r$^2$SCAN closely mimics the  SCAN functional. It is a good efficient alternative to SCAN due to its numerical efficiency.
In all cases, we observed that LSIC reduces the amount of SIC from PZSIC, but this reduction does not 
always improve the magnetic exchange couplings.
This is especially true for the systems that mainly consist of single electron regions where PZSIC already performs well.
For the more complex organic systems and [Cu$_2$Cl$_6$]$^{2-}$,
an overcorrecting nature of PZSIC is more pronounced when the SP approach is considered. 
We find that removing excess SIC using 
LSIC($z$) gives an improved performance over PZSIC in these cases.

In a case-by-case performance, combinations of certain functionals and SP/NP approaches work very well for a set of systems but work rather poorly for others. This includes 
$J_{SP}$ of PZSIC-PBE, -SCAN, and -r$^2$SCAN on H-He-H, $J_{NP}$ of  PZSIC-PBE and $J_{SP}$ of OSIC($w$)-LSDA on H$\cdots$He multicenter complexes, 
$J_{NP}$ of PZSIC-LSDA and PZSIC-PBE on organic molecules, and SCAN and r$^2$SCAN on chrolocuprate. 
Thus, some care must be exerted prior to the practical application of these methods for the evaluation of magnetic exchange couplings. 
On the other hand, we find that the SP approach with LSIC($z$)-LSDA provides a decently good overall performance across 
all four sets of systems, ranging from above average to a few of the best performances. This singles out the LSIC($z$) method as a promising option for applications to exchange coupling constants in a wide range of systems.  The performance of LSIC method can possibly be improved further by identifying more suitable iso-orbital indicator.

\begin{suppinfo}
\begin{itemize}
\item The equations used to determine the coupling constants of the H$\cdots$He multicenter complexes
\end{itemize}
\end{suppinfo}

\begin{acknowledgement}
Authors acknowledge Drs. Luis Basurto and Carlos Diaz for discussions and 
technical support and  
Prof. Mark R. Pederson for comments on the manuscript.
This work was supported by the US Department of Energy, Office of Science, Office of Basic Energy Sciences, as part of the Computational Chemical Sciences Program under Award No. DE-SC0018331. 
Support for computational time at the Texas Advanced 
Computing Center
and at NERSC is gratefully acknowledged.
\end{acknowledgement}

\bibliography{master}
\end{document}